\documentclass[preprint,12pt,authoryear]{elsarticle}

\usepackage{graphicx} 
\usepackage{amsmath}
\usepackage{amsfonts}
\usepackage{amssymb}
\usepackage{amsthm}
\usepackage{mathrsfs}
\usepackage{yfonts}
\usepackage{makecell}
\usepackage{afterpage}
\usepackage{printlen}
\usepackage{doi}

\usepackage{subcaption}
\DeclareCaptionSubType*[Alph]{table}
\DeclareCaptionLabelFormat{mystyle}{Table~\bothIfFirst{#1}{ }#2}
\usepackage[round]{natbib}
\graphicspath{{}}

\newtheorem{lemma}{Lemma}
\usepackage[a4paper, total={6in, 8in}]{geometry}

\begin{document}
\author{Ofek Aloni, Gal Perelman, Barak Fishbain}
\title{Synthetic Random Environmental Time Series Generation with Similarity Control, Preserving Original Signal's Statistical Characteristics}

\maketitle{}
\section*{Abstract}
Synthetic datasets are widely used in many applications, such as missing data imputation, examining non-stationary scenarios, in simulations, training data-driven models,  and analyzing system robustness. Typically,  synthetic data are based on historical data obtained from the observed system. The data needs to represent a specific behavior of the system, yet be new and diverse enough so that the system is challenged with a broad range of inputs. This paper presents a method, based on discrete Fourier transform, for generating synthetic time series with similar statistical moments for any given signal. The suggested method makes it possible to control the level of similarity between the given signal and the generated synthetic signals. Proof shows analytically that this method preserves the first two statistical moments of the input signal, and its autocorrelation function. The method is compared to known methods, ARMA, GAN, and CoSMoS. A large variety of environmental datasets with different temporal resolutions, and from different domains are used, testing the generality and flexibility of the method. A Python library implementing this method is made available as open-source software. 

\section*{Keywords}
Synthetic Data Generation, Environmental Simulations, Fourier Transform, Time Series Analysis, Urban water demand, Air Pollution, Wind analysis, Sea waves

\section{Introduction}
The development of computational models to analyze complex environmental and infrastructure systems depends to a great extent on the availability of datasets with sufficient size and quality. In many cases, these kinds of datasets are not available, so that synthetic data generators that can accurately mimic the properties of the real data are crucially needed. This is true for a wide range of environmental problems such as studies of air quality \citep{spec_air_qual}, wind speed \citep{synth_wind}, temporal and spatial analysis of fires \citep{synth_fire}, and weather characterization \citep{weather_gen}. Synthetic data are called for in cases of data unavailability, such as poorly recorded water demands time series, or too short historical datasets, that are not sufficient for computational methods \citep{nowak_2010}. Synthetic data are also commonly used for gap-filling, where only partial measurements are available \citep{kofinas_2018}. Another use of synthetic data is for examination of non-stationary scenarios which extend beyond the range of the historical records \citep{herman_synthetic_2016}. Even when data is available and stationary, a synthetic data generator is very useful for the analysis and optimization of models, since it can be harnessed to evaluate system behavior and robustness in response to different and broad ranges of scenarios.

To give a more concrete example, robustness of water resource systems has recently attracted a great deal of attention. Although there are many definitions of robustness in the literature, most concur that it reflects system performance over a set of future scenarios \citep{mcphail_2021}. Thus, in order to quantify robustness, a wide range of scenarios need to be examined. These scenarios should cover the range of values that have been observed in the past as well as future scenarios that include extreme values.

Generally speaking, synthetic data should comply with two requirements: the data should reflect the nature of the known parameters of historical records (time series) and at the same time allow for the manipulation of certain features to model non-stationary, less expected scenarios \citep{planning_water_2014}. In physical systems, the first requirement is especially crucial, as the simulated system is bounded by the laws of physics. To illustrate, one can think of a water distribution system, where the STD of water demands affects the the hourly distribution of the demands. Synthetic scenarios with different STDs might not reflect a reasonable behavior such that peak flows are in the middle of the night. Another example is the creation of water hammers due to abrupt changes in the synthetic flows that might effect the system dramatically. In ambient simulations, drastic changes in humidity or temperature are simply not feasible.

Many studies have tackled the challenge of developing a generalized synthetic data generator. Early approaches included parametric methods such as Auto Regressive (AR) models and the Auto Regressive Moving Average (ARMA and ARIMA) models \citep{box1990}, which are still common today. For example, \cite{papamichail2001} generated a synthetic monthly reservoir inflow series using the Seasonal ARIMA (SARIMA) model. \cite{talbot2020} coupled ARMA with a Fourier series to generate electricity grid demands.

Non-parametric approaches for synthetic data generation are based on resampling and reshuffling historical data values. One of the most common non-parametric approaches is bootstrapping \citep{efron1992}. This includes the Moving Block Bootstrap (MBB), the Circular Block Bootstrap (CBB), and the Stationary Bootstrap (SB). To take into account the original signal's autocorrelation, the bootstrap resampling is carried out in blocks of consecutively sampled values (\cite{hirsch2015}; \cite{selle2010}). However, block bootstrapping cannot handle the correlation between blocks. Another type of bootstrap method is the K Nearest Neighbor bootstrap (K-NN) which does maintain the block correlation \citep{nowak_2010}. However, all bootstrap methods are based on resampling the input data records. This ensures almost exactly the same statistical properties as the input, but causes the synthetic dataset to consist only of values found in the historical signals, and hence the generated time series has no new values.

A more recent approach for generating synthetic time series is based on data-driven models, such as neural networks. To this end, NN based approaches have been presented for water demand \citep{odan2012} and water quality forecasting \citep{chen2020}. A recent data-driven approach for synthetic data generation is the Generative Adversarial Network (GAN). GAN models were first introduced by \cite{GAN_og_paper}  as a tool to generate synthetic data for synthetic images. This concept was extended to a time-series synthetic generator by \cite{yoon2019} who demonstrated its performance on stock price data, energy consumption, and health related acute lung emergencies. 

Stochastic models can also be a powerful tool for synthetic data generation. For example, in generating climate data \citep{wilks1999, srikanthan2001, efstratiadis2014}. Such models can be difficult to generalize, if they are tailored for a specific type of data. For example, precipitation might have a very different probability distribution than temperature measurements. To attempt to resolve this issue, \cite{papalexiou2018} presents a framework for stochastic generation of univariate and multivariate hydroclimatic data, while \cite{papalexiou2020} has a similar goal with spatiotemporal random fields.

A bottom-up approach, where the analysis starts at the end node, and then is aggregated to a whole system level, has been introduced for the simulation of demand pulse distributions in water distribution systems \citep{buchberger1995}. Improvements to the pulse-based approach were presented by \cite{blokker2010} and \cite{creaco2016}. Another bottom-up approach was presented by \cite{alvisi2014}, where high-resolution data (every 1 minute) per consumer were synthesized, and then aggregated over temporal and spatial axes. This approach preserved the mean, the variance, and the temporal covariance of the historical records but required detailed knowledge about the consumers to enable the high-resolution data synthesis. \par

When solving optimization problems related to the design and operation of complex environmental systems, the aggregated behavior is more important than the single user or end-node. For this reason, studies have also suggested top-down methods for the generation of synthetic time series. One example of a top-down approach presented by \cite{brentan2018} used a Random-Forest algorithm for generating a random data series based on historical data combined with climatic features. \cite{kossieris2019} suggested a stochastic approach combined with Nataf's joint distribution model to simulate water demands at different time scales. \cite{santopietro2020} used Hermite interpolating polynomials coupled with a mixed distribution approach to generate a random data series that met known values of the daily trend.

Clearly, there are many existing methods for synthetic signal generation. However, these methods do not necessarily preserve the mean, standard deviation, and autocorrelation of the original series. Often they are domain specific, especially when data characteristics such as stationarity, seasonality, sampling scale, or series length need to be accounted for. Additionally, in many of the above methods a significant amount of data is required as input.

Here, we present a simple, easy to implement, Fourier based synthetic time series generator, that preserves the signal's first two statistical moments and the autocorrelation function. It can take any signal as input, from any domain, and it allows the user to select the level of similarity, in terms of dynamic time wrapping or wasserstein distance \citep{similarity_eval}, between the original time series and the synthesized signals. The discrete Fourier transform is a widely used mathematical tool, and efficient algorithms to execute it exist in widespread programming languages. This makes the generation method presented here quicker than many of the alternatives, which can be especially meaningful when working with large datasets, or producing a large batch of samples. 

\section{Methodology}

\subsection{Synthetic Signal Generator} \label{genmethod}

Let $S=\{S_k\}_{k=0}^{N-1}$ be a discrete time-series signal with length $N$, mean $\mu_S$, and standard deviation $\sigma_S$. The Discrete Fourier Transform (DFT) of $S$, denoted $\zeta_\omega$ for $\omega = 0,\dots,N-1 $ by:
\begin{equation}
\zeta_\omega = \mathcal{F}\{S\}_\omega := \sum\limits_{k=0}^{N-1} S_k e^{-i\frac{2\pi  k}{N}\omega}
\end{equation}

Since $\zeta_\omega$ is a complex signal, we can use a polar representation:

\begin{equation}
\zeta_\omega = \rho_\omega e^{i \theta_\omega}
\end{equation}

Where $ 0 \leq \rho_{\omega} \in \mathbb{R}$ is the amplitude and $\theta_\omega \in \mathbb{R}$ is the phase.

The generation of a synthetic signal $\hat{\zeta}_\omega$ with statistical properties of a given base signal involves the manipulation of the polar representation. Specifically, it consists of replacing the phase terms $\theta_\omega$ with random phase values $\hat{\theta}_\omega$. This makes it possible to get a new random time series and keep the original signal energy because the amplitudes of the decomposed signals remain unchanged. To guarantee the conservation of the mean, the zero phase is kept unchanged, $\hat{\theta_0} = \theta_0
$. In addition to this phase, one may choose to keep  $0<m \leq N$ of the original phases, meaning $\hat{\theta_j} = \theta_j$ for any $0<j\leq m$. This parameter allows for adjusting the synthetic signal's similarity to the original signal, as is detailed below.

After randomizing the phase components, a synthetic signal, $\hat{S}_k$, is generated by applying the inverse DFT:
\begin{equation}
    \hat{S_k} =\mathcal{F}^{-1}\{\hat{\zeta}\}_k := \frac{1}{N}\sum\limits_{\omega=0}^{N-1} \hat{\zeta}_\omega e^{i \frac{2\pi  \omega}{N}k} 
\end{equation}
This gives us a synthetic time series $\hat{S}$ of the same length, sampling rate, and the first two statistical moments as the original time series.

One issue with this method is that the inverse transform is not guaranteed to consist only of real values. To rectify this we exploit the mathematical properties of the DFT of a real signal, so we can ensure that the synthetic signal is also real. To this end, let us consider a real signal $\{S_k\}\subset \mathbb{R}$. Its DFT holds the following conjugate symmetry:

\begin{align}
\zeta_{\omega} = \overline{\zeta}_{N-\omega}, && 1 \leq \omega \leq N-1
    \end{align}

This is equivalent to the following choices of $\hat{\theta}$ , providing that we identify $-\pi$ with $\pi$:

\begin{align}
\hat{\theta}_{N-\omega} = -\hat{\theta}_{\omega} , && 1 \leq \omega \leq N-1
\end{align}

Note that if $N$ is even, the conjugate symmetry has a special implication for $\zeta_{N/2}$:
\begin{equation}
    \zeta_{N/2} = \bar{\zeta}_{N-N/2} = \bar{\zeta}_{N/2}
\end{equation}
And it is it's own conjugate, meaning $\zeta_{N/2}\in \mathbb{R}$ . Thus, in order to ensure that $\hat{S}$ is real, after the randomization process in the Fourier domain, $\hat{\zeta}_{N/2}$ must fulfill:
\begin{equation}
    \hat{\zeta}_{N/2} = \pm \zeta_{N/2}
\end{equation}
Practically speaking,the suggested methodology is easy to implement, as many libraries implement DFT and its inverse. The constraints above on the phase randomization process ensures that the inverse tranform will produce a real series. The only element that requires special consideration, in the case of a time series with an even number of elements, is  $\hat{\zeta}_{N/2}$.

Thus, for a given number of phases $0<m\leq \frac{N}{2}$ the formulation for phase randomization is as follows. For the even $N$ case:

\begin{equation}
\begin{cases}
    \hat{\theta}_\omega = \theta_\omega, & \omega=0,\dots,m\\
    \hat{\theta}_\omega = rand(-\pi, \pi], & \omega = m+1, \dots, \frac{N}{2}-1\\
    \hat{\theta}_{N/2} = 0 \text{ or } \pi\\
\end{cases}
\label{even_rand}
\end{equation}
And for the odd case:
\begin{equation}
\begin{cases}
    \hat{\theta}_\omega = \theta_\omega, & \omega=0,\dots,m\\
    \hat{\theta}_\omega = rand(-\pi, \pi], & \omega = m+1, \dots, \frac{N-1}{2}\\
\end{cases}
\label{odd_rand}
\end{equation}

The parameter $m$ defines how many of the first components will preserve their phase, allowing to control the similarity between the original signal $S$ and the synthetic signal $\hat{S}$. As we increase the number of phases that are not randomized, we increase the similarity of the decomposed periodic components until we reach the point where the two signals are identical in the case of $m=\frac{N}{2}$. This capability of controlling the similarity level of the synthetic signal and the original one, while preserving the signal's statistical moments, is unique to this method.

A special case to consider is a synthetic series that is inherently non-negative, but with typical values that are close to zero.  Using the randomization process described above is likely to generate synthetic series with negative values, which might not be acceptable for certain cases. As a way to amend this issue, the following optional procedure is offered. Taking the synthetic series $\hat{S}$ as our original series, a new synthetic series $\hat{\hat{S}}$ is generated using the same method. Any negative values of $\hat{S}$ are replaced with the corresponding values of $\hat{\hat{S}}$ (that are not guaranteed to be non-negative). If any negative values remain in $\hat{S}$ , this process is repeated and a new $\hat{\hat{S}}$ is generated. This is iterated until all values in $\hat{S}$ are non-negative. It is important to note that this process alters the statistical properties of $\hat{S}$, and while using it the mean and SD are not conserved.

\subsection{Proofs of Conservation of Mean, Standard Deviation, and Auto-correlation}
\begin{lemma}
The mean of \(S\) is equal to the mean of \(\hat{S}\),  \(\mu_S=\mu_{\hat{S}} \).
\end{lemma}
\begin{proof} From the definition of \(\hat{S}\):
\[\hat{S}_k = \mathcal{F}^{-1} \{\zeta_{\omega}\} = \frac{1}{N} \sum^{N-1}_{\omega=0} \hat{\zeta}_\omega e^{-i\frac{2\pi}{N} k \omega}\]
Plugging it into the definition of the mean:
\[ \mu_{\hat{S}}=\frac{1}{N} \sum^{N-1}_{k=0}\hat{S}_k  
 = \frac{1}{N^2} \sum^{N-1}_{k=0} \sum^{N-1}_{\omega=0} \hat{\zeta}_\omega e^{-i\frac{2\pi}{N} k \omega} = \frac{1}{N^2}( N \cdot\hat{\zeta}_0 + \sum^{N-1}_{k=0} \sum^{N-1}_{\omega=1} \hat{\zeta}_\omega e^{-i\frac{2\pi}{N} k \omega}) \]
 We shall now show that the sum $\sum^{N-1}_{k=0} \sum^{N-1}_{\omega=1} \hat{\zeta}_\omega e^{-i\frac{2\pi}{N} k \omega}$ vanishes. Since this is a finite sum, we can flip the order of summation, and then use the formula for a sum of a geometric series:
 \[\sum^{N-1}_{k=0} \sum^{N-1}_{\omega=1} \hat{\zeta}_\omega e^{-i\frac{2\pi}{N} k \omega} 
 =  \sum^{N-1}_{\omega=1} \hat{\zeta}_\omega \sum^{N-1}_{k=0}  e^{-i\frac{2\pi}{N} k \omega}
 = \sum^{N-1}_{\omega=1} \hat{\zeta}_\omega \cdot \frac{1-\overbrace{e^{-i\frac{2\pi}{N} \cdot N \omega}}^{=1}}{1-e^{-i\frac{2\pi}{N}\omega}} = 0\]
 
Since the first transform coefficient is retained, \(\hat{\zeta_{0}}=\zeta_0\), we get:
\[\hat{\zeta}_0 = \zeta_0 = \sum^{N-1}_{k=0} S_k \underbrace{e^{-i\frac{2\pi}{N}k\cdot 0}}_{=1} = \sum^{N-1}_{k=0} S_k\]
And so:
\[ \mu_{\hat{S}} = \frac{1}{N}\cdot\hat{\zeta}_0 = \frac{1}{N} \sum^{N-1}_{k=0} S_k = \mu_S \] 

\end{proof}

\begin{lemma}
    The standard deviation of \(S\) is equal to the standard deviation of $\hat{S}$, \( \sigma_S = \sigma_{\hat{S}}\).
\end{lemma}
\begin{proof}
    Writing out the definition for discrete standard deviation, and expanding, we get:
    \[
     \sigma^2_{\hat{S}}=\frac{1}{N} \sum_{k=0}^{N-1}\left(\hat{S}_k-\mu_{\hat{S}}\right)^2=\frac{1}{N} \sum_{k=0}^{N-1} (\hat{S}_k^2- 2 \hat{S}_k \mu_{\hat{S}}+\mu_{\hat{S}}^2)
     =  \sum_{k=0}^{N-1} \hat{S}_k^2 - 2 \mu_{\hat{S}}^2 +  \mu_{\hat{S}}^2 
     = \frac{1}{N} \sum_{k=0}^{N-1}\left|\hat{S}_k\right|^2-\mu_{\hat{S}}^2
    \]
    From lemma 1, we know \(\mu_S=\mu_{\hat{S}}\). From Parseval's Theorem, we get:
    \[
    \sum_{k=0}^{N-1}\left|\hat{S}_k\right|^2 = \frac{1}{N} \sum^{N-1}_{\omega=0}|\hat{\zeta}_\omega|^2
    \]
    Since the amplitudes are not changed during the process, \(\rho_\omega=\hat{\rho}_\omega\), we get:
    \[
    |\hat{\zeta}_\omega|^2 = |\zeta_\omega|^2
    \]
    And so applying the identity from Parseval's theorem twice:
    \[
    \sum_{k=0}^{N-1}\left|\hat{S}_k\right|^2=\frac{1}{N} \sum_{\omega=0}^{N-1}\left|\hat{\zeta}_\omega\right|^2=\frac{1}{N} \sum_{\omega=0}^{N-1}\left|\zeta_\omega\right|^2=\sum_{k=0}^{N-1}\left|S_k\right|^2
    \]
    And so:
    \[
    \sigma^2_{\hat{S}}
    = \frac{1}{N} \sum_{k=0}^{N-1}\left|\hat{S}_k\right|^2-\mu_{\hat{S}}^2
    = \frac{1}{N} \sum_{k=0}^{N-1}\left|S_k\right|^2-\mu_{S}^2
    = \sigma^2_S
    \]
    \[
    \Rightarrow \sigma_{\hat{S}} = \sigma_S 
    \]
\end{proof}

\begin{lemma}
       $S$ and \(\hat{S}\) have the same autocorrelation functions.
\end{lemma}
\begin{proof}
    As known by the Wiener-Khinchin Theorem,
    \[
    Autocorr(S) = \mathcal{F}^{-1} \{|\zeta|^2\}
    \]
    As mentioned before, since the amplitudes are unchanged, \(|\hat{\zeta}_\omega|^2 = |\zeta_\omega|^2\), and so:
    \[
    Autocorr(\hat{S}) = \mathcal{F}^{-1} \{|\hat{\zeta}|^2\} = \mathcal{F}^{-1} \{|\zeta|^2\} = Autocorr(S) 
    \]
\end{proof}

\subsection{Comparison}
The above characteristics of the Fourier based method make it easy and convinient to use. However, many synthetic signal generation methods have been presented in the literature. Here, we use the following for comparison.

\subsubsection{Auto Regressive Moving Average (ARMA)}

The first method used for comparison is the Auto Regressive Moving Average (ARMA). While ARMA is typically used for time series forecasting, it can also be used for generation. It combines an autoregressive model $AR(p)$, with a moving average model $MA(q)$. Briefly, in an $ARMA(p,q)$ model a given time step $y_t$ is calculated using the $p$ previous time steps, and $q$ past noise elements $\varepsilon_t$, as such:
\begin{equation}
y_t=c+\phi_1 y_{t-1}+\cdots+\phi_p y_{t-p}+\theta_1 \varepsilon_{t-1}+\cdots+\theta_q \varepsilon_{t-q}+\varepsilon_t
\end{equation}
Where $c, \phi_{1,\dots,p}, \theta_{1,\dots,q}$ are the model's parameters. See \cite{fpp3} for more details.
In our implementation we used the Statsmodels python library \citep{statsmodels_ref} to find the model's parameters given the hyperparameters $p,q$, and then generate time series accordingly.

\subsubsection{GAN}

As mentioned before, a much newer method for signal generation is Generative Adversarial Networks (GAN). The general idea is to place two neural networks in a zero sum game. The first model, the aptly named Generator, is tasked with generating data. The second model, the Discriminator, is attempting to distinguish between the synthetic data and the original data set. The models are trained side by side, with the generator training to "fool" the discriminator, which in turn trains to improve at discerning fake data. A trained model can generate a lot of synthetic data relatively quickly, and if the process converged properly it should resemble the original dataset. It is important to note that a major characteristic of GAN is that it acts as black box - no direct control of the generator model is possible before or after training.
The architecture used for comparison is quantGAN \citep{quantGAN}, designed for modelling financial time series. Although intended for a different field, it fits our use case. Using temporal convolutional networks, it can capture both the distribution of the data, and the autocorrelation in different lags. This is in contrast to GANs intended for tabular data \citep{CTGAN}, that emulate the distribution but not the temporal dependencies. QuantGAN accepts a single series as input of any length, allowing us to train it on each dataset used for comparison. The hyperparamters for each dataset were determined by trial and error.

The main deviation from the methodology of the original quantGAN is in the way our data was scaled in pre-processing. For proper performance, the time series values fed into the GAN should be in the range $(-1,1)$. In the original paper, the scaling is done in a method that relies on specific properties of the financial data used. To achieve this for any time series, the MinMaxScaler function from Scikit-learn \citep{scikit} was used, as it gave the best results of all scalers tested. The original data was transformed in pre-processing, and the generated series were transformed back to the original scale using the inverse operation. 

\subsubsection{CoSMoS}
CoSMoS \citep{cosmos_R_package} is an R package for stochastic generation of synthetic data, with a focus on hydroclimatic datasets. One use case is generating synthetic data from scratch, by choosing a probability distribution for the data, and an autocorrelation structure (ACS). However, it can also take an existing time series. One chooses the type of distribution appropriate for the data type (Generalized Gamma, Burr, etc.) and the type of ACS. Then the parameters of the distribution and ACS are fitted to the given time series. This allows generation of new synthetic time series with the same probability distribution and ACS as the original.

\subsection{Data}
A variety of environmental datasets were chosen as examples for this paper. This was done to offer varying conditions for testing the performance of the suggested method. The data consists of domestic water demand, concentration of air pollutants, wave heights, and turbulent air flow velocities. These signals differ in their length, the rate of change (i.e. frequency components) and the typical observed values. Some are inherently positive and fluctuate close to zero, while others change sign often or have large positive values throughout.

For each dataset, 1,000 synthetic series were generated by each method. This was done since all generation methods have a random element, so analyzing the signal's properties over a large set provides a more accurate insight.

\subsubsection{Urban Water Demand}
The urban water demand dataset contains hourly values of water demand of a metropolitan in cubic meters, taken from the connection between the metro and the national water company. The metro has about 1 million residents, and is spanning over 126 squared kilometers. Average height above sea level is 715 meters. Temperature averages are $25^{\circ}$ Celsius in the hottest months, July-August, and about $10^\circ$ in the coldest, January. The full data spans a year, from 1/1/2020 until 31/12/2020. For clearer data visualization, only the month of July was used, giving 745 hourly time-steps. 

\subsubsection{Air Pollution}
Dataset containing urban ambient nitric oxide (NO) levels, in $[\frac{\mu g}{m^3}]$, as measured by a standard monitoring station, was obtained. The station was positioned 20 meters above ground, and 235 meters above sea level. The city has approximately 300,000 residents, over an area of about 63.5 squared kilometers. The data contains daily averages, from 1/1/2023 to 31/12/2023. 
An additional example of air pollution dat, of Fine Particulate Matter (PM 2.5) measurements from the same station, exists in the supplementary material.

\subsubsection{Air Flow Velocity}
Turbulent Air flow velocity, measured in an arbitrary location near Nofit, 18 km east of Haifa, Israel, taken 179 meters above sea level.
The original files contain three orthogonal velocity field components $\vec{V}= (u(t),v(t),w(t))$, measured in $[m/s]$. The measurements sampling frequency was 2 kHz, and one minute of measurement was used. The $u$ component was arbitrarily selected to represent this set of measurements.
 
\subsubsection{Wave Height}
Another enviromental domain used here is wave field synthetic data generation. To this end, wave height records from the gulf of Eilat were used  \citep{shani-zerbib2017,shani-zerbib2018}. Data was collected continuously for over 50 hours during June 2017, taken with an array of 5 wavegauges, positioned in a circle. Measurements are statistically identical between the gauges, with variations in phase.  Sampling frequency was 80Hz. The full set of measurements is available via Mendeley \citep{wave_height_data}.

\subsection{Comparison of Synthetic Time Series}
Quantitative methods of comparison between the generated time series are essential to understanding the differences between the generation methods. 4 statistical measures were chosen - mean, standard deviation (SD), skewness, and kurtosis. The values were calculated for each synthetic series, then averaged over all 1,000 samples.

However, statistical properties lack the ability to reflect seasonal trends, and to describe how similar are a pair of compared time series in any specific point in time. A pair of time series may have a similar mean and SD, and still have vastly different plots and distributions. Thus additional similarity measures, specific to time series analysis were used. Here, two metrics highlighted in \cite{GAN_ts_review} were employed - Dynamic Time Warping and the Wasserstein Distance.
 
Dynamic Time Warping (DTW) measures of the distance between a pair of time series. DTW and its variants are often used for time series similarity analysis, and are considered to be a good metric for comparison \citep{similarity_eval}. Given two time series $X=\{x_i\}^n_{i=1}$ and  $Y=\{y_j\}^m_{j=1}$, one can create a path $\pi$ between them, matching each element in $X$ to an element of $Y$, and vice versa. This matching is not necessarily one-to-one, but all points should have a match. In addition, $x_1$ is always matched to $y_1$, and $x_n$ to $y_m$ . The matching needs to respect order (the path cannot "cross" itself). Then DTW is computed as a minimum of the Euclidean distance on all such paths:
\begin{equation}
    D T W(x, y)=\min _\pi \sqrt{\sum_{(i, j) \in \pi} d\left(x_i, y_j\right)^2}
\end{equation}
We compare all synthetic series generated to the original using DTW, and then take the average and standard deviation (SD). The \textbf{lowest} score represents the method generating synthetic signals that are the most similar to the original series. 

Wasserstein distance (WD, also known as Earth Mover distance), measures the "cost" of turning one distribution into another, giving a way to quantitatively compare distributions. Here we refer to Wasserstein-1. Given probability mass functions $u,v$ (corresponding to time series, in our case) the formula is:
\begin{equation}
    W_1(u, v)=\inf _{\pi \in \Gamma(u, v)} \int_{\mathbb{R} \times \mathbb{R}}|x-y| \mathrm{d} \pi(x, y)
\end{equation}
Where  $\Gamma (u,v)$ are the set of all joint probability measures on $\mathbb{R} \times \mathbb{R}$ with marginal distributions $u,v$, respectively. Here as well, we compare all synthetic series to the original, then take the average and SD of the results. The \textbf{lowest} WD score belongs to the method that generates series with distributions that are the most similar to the original distribution. 

Synthetic data should maintain characteristics of the original data it is trying to replicate, so some similarity to the original data is desirable. However, if it is too similar to the original data, it limits the range of possible outcomes considered, and thus it will not be able to predict cases differing substantially from the given original sample. Hence the optimal similarity level is application dependant. The method in hand, as described above, presents a mechanism to control the similarity level.

Two other useful tools for time series analysis are the distribution function, and the Autocorrelation Function (ACF) plot. Those were also produced for a single series from each generation method for every dataset, to give more insight about the temporal dependencies.

\section{Results \& Discussion}
\subsection{Control Similarity}
\begin{figure}[p]
    \centering
    \includegraphics[width=1\linewidth]{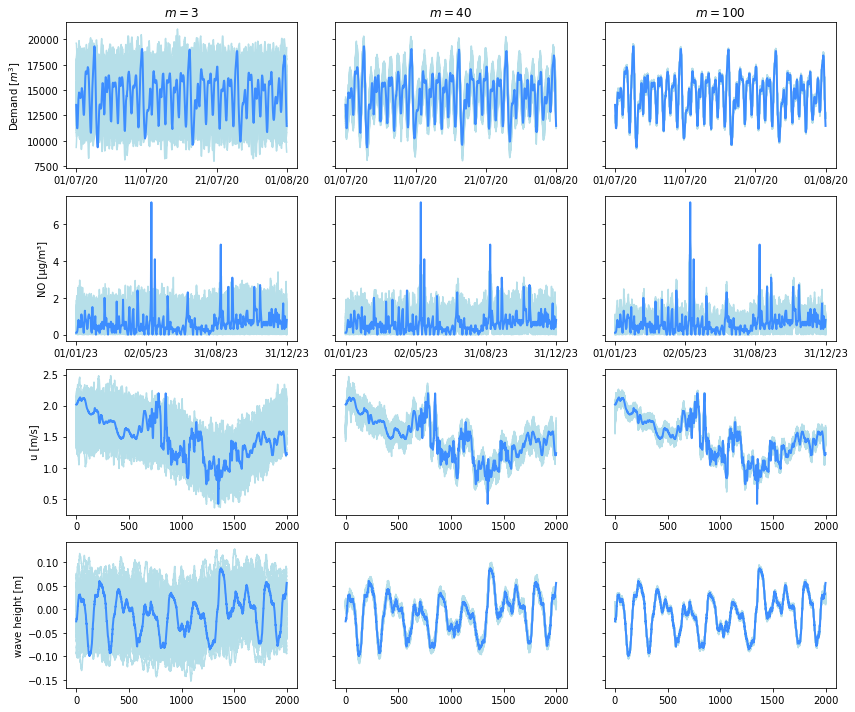}
    \caption{Comparing different $m$ values for each dataset. }
    \label{fig:m_comparison}
\end{figure}

One of the key features of the suggested method is the ability to control the synthetic generated signal's similarity to the original signal. This is done through the number of phase components that are randomized (see Eq. \ref{even_rand} and Eq. \ref{odd_rand}). The effect of randomized components is illustrated in Figure \ref{fig:m_comparison}, where the original signal and a cohort of synthetic signals are presented for all datasets in this study. The original signal is presented in blue, and in the background a set of 100 synthetic signals are plotted in light blue. The left column in Figure \ref{fig:m_comparison} depicts the original signal in front of a synthesized set, where only the first three phase components are preserved, while the rest are randomized. The second column presents the same configuration with 40 preserved phase coefficients. The right-hand column plots the original signal on top of 100 synthetic signals that have been generated with 100 preserved phase coefficients. It is clear that the more coefficients are preserved, the higher the similarity of the generated signals. This is also illustrated in table \ref{m-comparison} . In particular, from table \ref{m-comparison-metrics} and Figs. \ref{fig:ud_comparison}-\ref{fig:wave_height_comparison} we can discern that the similarity between the original and the synthetic signals increases with $m$, as evident by the decrease in DTW and WD. Yet, for any $m$ value, the mean and SD are preserved.

\begin{table}[hbp!]
    \centering
    \caption{Statistics and metrics averaged over 100 samples for each specified $m$ value, using the Fourier method.}
    \label{m-comparison}
    \begin{subtable}{\textwidth}
            \centering
        \caption{Mean, Standard Deviation (SD), Skewness, and Kurtosis for each $m$ value.}
        \label{m-comparison-stats}
\resizebox{\linewidth}{!}{%
\begin{tabular}{|c|c|c|c|c|}
\hline
                                       & \textbf{Mean}     & \textbf{SD}      & \textbf{Skewness} & \textbf{Kurtosis} \\ \hline
\textbf{\makecell{Water Demand $[m^3]$}}& 14528             & 1915             & -0.26             & -0.34             \\ \hline
$m=3$& 14528 $\pm$ 0& 1916 $\pm$ 0& -0.02 $\pm$ 0.31     & -0.43 $\pm$ 0.25     \\ \hline
$m=40$& 14528 $\pm$ 0& 1916 $\pm$ 0& -0.09 $\pm$ 0.28     & -0.20 $\pm$ 0.33     \\ \hline
$m=100$& 14528 $\pm$ 0& 1916 $\pm$ 0& -0.25 $\pm$ 0        & -0.35 $\pm$ 0.02     \\ \hline
\textbf{\makecell{NO {[}µg/m³{]}}}& 0.64              & 0.65             & 4.58              & 34.25             \\ \hline
$m=3$& 0.75 $\pm$ 0.01& 0.52 $\pm$ 0.01& 0.71 $\pm$ 0.12& 0.08 $\pm$ 0.41\\ \hline
$m=40$& 0.74 $\pm$ 0.01& 0.54 $\pm$ 0.01& 0.92 $\pm$ 0.13& 0.77 $\pm$ 0.6\\ \hline
$m=100$& 0.7 $\pm$ 0.01& 0.59 $\pm$ 0.01& 1.97 $\pm$ 0.17& 6.8 $\pm$ 1.5\\ \hline
\textbf{\makecell{u {[}m/s{]}}}& 1.47              & 0.33             & 0.003             & -0.56             \\ \hline
$m=3$& 1.47 $\pm$ 0& 0.33 $\pm$ 0& -0.18 $\pm$ 0.22     & -0.50 $\pm$ 0.35     \\ \hline
$m=40$& 1.47 $\pm$ 0& 0.33 $\pm$ 0& 0.02 $\pm$ 0.02      & -0.55 $\pm$ 0.06     \\ \hline
$m=100$& 1.47 $\pm$ 0& 0.33 $\pm$ 0& 0.02 $\pm$ 0.01      & -0.61 $\pm$ 0.01     \\ \hline
\textbf{\makecell{Wave  Height {[}m{]}}}& -0.0092           & 0.04             & -0.17             & -0.63             \\ \hline
$m=3$& -0.0092 $\pm$ 0& 0.04 $\pm$ 0& -0.009 $\pm$ 0.272& -0.273 $\pm$ 0.414\\ \hline
$m=40$& -0.0092 $\pm$ 0& 0.04 $\pm$ 0& -0.182 $\pm$ 0.007& -0.576 $\pm$ 0.026\\ \hline
$m=100$& -0.0092 $\pm$ 0& 0.04 $\pm$ 0& -0.172 $\pm$ 0.001& -0.621 $\pm$ 0.002\\ \hline
\end{tabular}
}
\end{subtable}
\end{table}

\pagebreak

\begin{table}
\ContinuedFloat
    \begin{subtable}{\textwidth}
            \centering
        \caption{Metric values for both Dynamic Time Warping (DTW) and Wasserstein Distance (WD), quantifying the similarity of the synthetic samples to the original time series for each $m$ value. Lower scores indicate greater similarity.}
        \label{m-comparison-metrics}
\begin{tabular}{|c|c|c|}
\hline
                                       & \textbf{DTW}     & \textbf{WD}      \\ \hline
\textbf{\makecell{Water Demand  $[m^3]$}}& -                & -                \\ \hline
$m=3$& 21866 $\pm$ 1868    & 230 $\pm$ 98        \\ \hline
$m=40$& 13033 $\pm$ 1766    & 239 $\pm$ 104       \\ \hline
$m=100$& 3621 $\pm$ 216      & 28 $\pm$ 2          \\ \hline
\textbf{\makecell{NO  {[}µg/m³{]}}}& -                & -                \\ \hline
$m=3$& 9.2 $\pm$ 0.21& 0.19 $\pm$ 0.01\\ \hline
$m=40$& 8.78 $\pm$ 0.23& 0.17 $\pm$ 0.01\\ \hline
$m=100$& 7.54 $\pm$ 0.38& 0.11 $\pm$ 0.01\\ \hline
\textbf{\makecell{u  {[}m/s{]}}}& -                & -                \\ \hline
$m=3$& 5.41 $\pm$ 0.81     & 0.03 $\pm$ 0.01     \\ \hline
$m=40$& 2.63 $\pm$ 0.15     & 0.01 $\pm$ 0.005\\ \hline
$m=100$& 1.65 $\pm$ 0.05     & 0.01 $\pm$ 0.005\\ \hline
\textbf{\makecell{Wave  Height  {[}m{]}}}& -                & -                \\ \hline
$m=3$& 0.687 $\pm$ 0.109& 0.005 $\pm$ 0.002\\ \hline
$m=40$& 0.171 $\pm$ 0.001& 0.001 $\pm$ 0.0001\\ \hline
$m=100$& 0.111 $\pm$ 0.003& 0.0004 $\pm$ 0.0001\\ \hline
\end{tabular}
\end{subtable}%
\end{table}

\afterpage{\clearpage}
\newpage

 \subsection{Urban Water Demand}
\begin{figure}[p]
    \centering
    \includegraphics[width=1\linewidth]{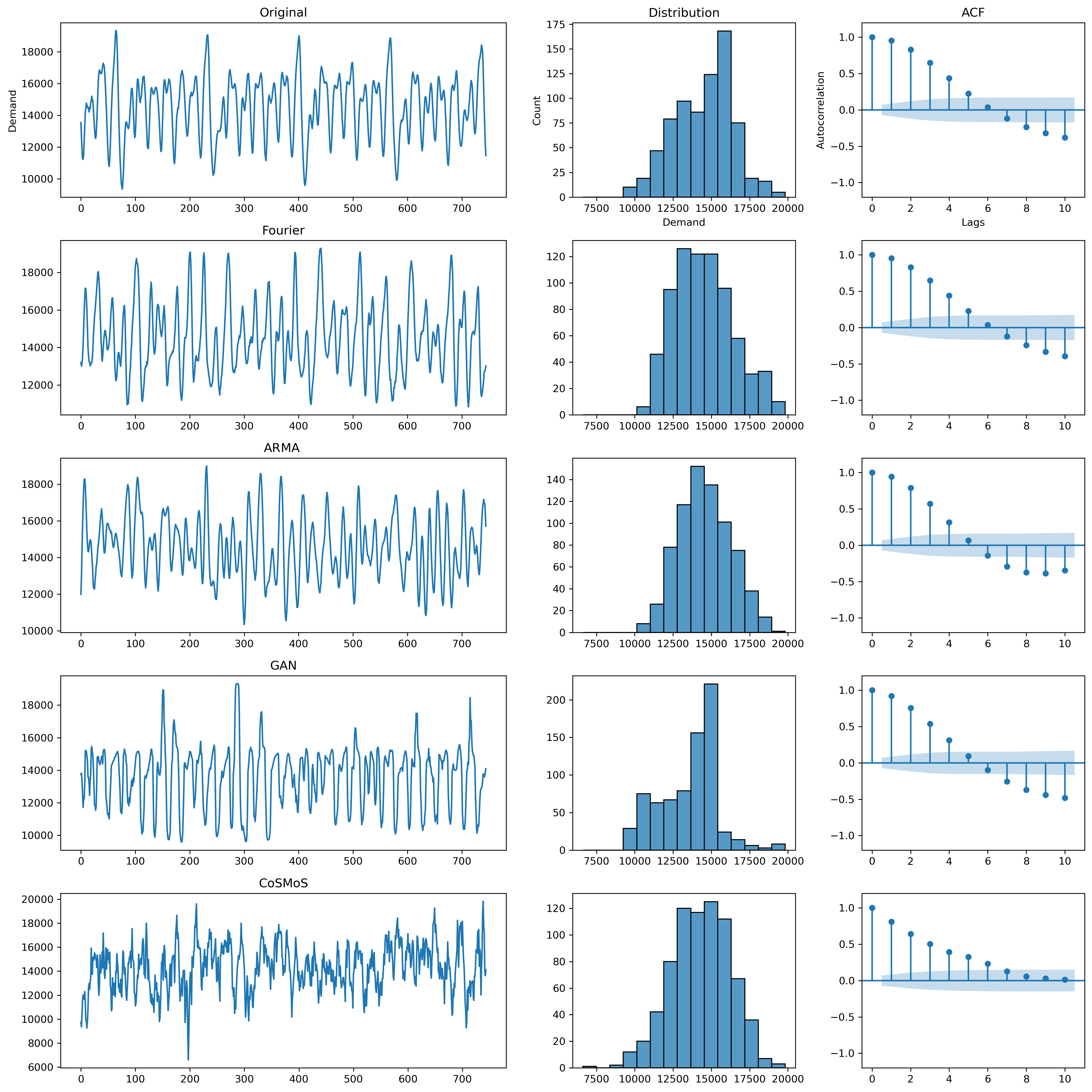}
    \caption{The original time series (at the top) and a single time series produced by each method for the Urban Demand dataset, including distribution and ACF for each. Colored area in ACF depicts the confidence interval of $95\%$ }
    \label{fig:ud_comparison}
\end{figure}

For urban water demand we used $m=3$ for the DFT method. The model parameters for ARMA were selected to be $p=3,q=3$, as those have provided the best results in terms of visual similarity to the original. For CoSMoS, the distribution was set to Generalized Gamma, and the ACS was Pareto II. As seen in Fig. \ref{fig:ud_comparison} , the autocorrelation of the original series contains negative correlations that cannot be obtained by Pareto II. This is the case for the included ACSs, and our attempts to create a custom ACS were unsuccessful.

Table \ref{tab:ud_stats} presents the mean, standard deviation (SD), skewness, kurtosis, DTW and WD measures for the urban water demand time series. Fourier retains the mean and SD, as expected. The skewness is not retained on average, but the Fourier method does produce the closest value to the skewness of the original series. Examining the SD of the skewness suggests that the individual synthetic series have similar skewness to the original, but with alternating signs.  The average kurtosis of the series generated by the Fourier method, presented here, is closest to the original. Fourier also achieved the lowest DTW score, $10\%$ lower than the next, ARMA, and $20\%$ lower than the highest, CoSMoS. Fourier was best in WD as well, $12\%$ lower than ARMA. Note however the large variability, as the SD is 42\% of the average score.

\begin{table}[ht]
    \centering
        \caption{Statistics and metrics for Urban Demand Dataset. Each value is the average of all 1,000 synthetic series produced with the same method, $\pm$ SD. In bold are the scores indicating greates similarity to the original series, meaning the closest mean, SD, Skewness and Kurtosis, and the lowest DTW and WD scores.}
\resizebox{\linewidth}{!}{%
\begin{tabular}{|c|c|c|c|c|c|} 
\hline
\textbf{m=3}& Original & Fourier                   & ARMA             & GAN              & CoSMoS            \\ 
\hline
Mean         & 14528    & \textbf{14528} $\pm$ 0    & 14536 $\pm$ 160  & 13645 $\pm$ 41   & 14526 $\pm$ 217   \\ 
\hline
SD           & 1916     & \textbf{1916} $\pm$ 0     & 1909 $\pm$ 123   & 1844 $\pm$ 161   & 1881 $\pm$ 108    \\ 
\hline
Skewness     & -0.26    & 0 $\pm$ \textbf{0.31}     & -0.00 $\pm$ 0.19 & -0.08 $\pm$ 0.17 & -0.04 $\pm$ 0.19  \\ 
\hline
Kurtosis     & -0.34    & \textbf{-0.43} $\pm$ 0.26 & -0.05 $\pm$ 0.32 & 0.11 $\pm$ 0.31  & 0.16 $\pm$ 0.45   \\ 
\hline
DTW          & -        & \textbf{22087} $\pm$ 1942 & 24518 $\pm$ 1783 & 26866 $\pm$ 1343 & 26894 $\pm$ 1796  \\ 
\hline
WD           & -        & \textbf{236} $\pm$ 100    & 266 $\pm$ 62     & 894 $\pm$ 42     & 275 $\pm$ 82      \\
\hline
\end{tabular}
}
\label{tab:ud_stats}
\end{table}

Fig. \ref{fig:ud_comparison} takes the original series, as well as a random synthetic series from each of the generation methods, and plots them together for comparison. A histogram is plotted, using the same bins for all the plots for easier assessment. The autocorrelation function (ACF) is plotted up to 10 lags. 
Looking at Fig. \ref{fig:ud_comparison}, GAN has a similar autocorrelation shape, and the plot is visually similar to the original. However, GAN recieved the lowest WD score, 3.8 times larger than Fourier, and the second-worst DTW score.  This disparity might be explained by the average mean, that is $6\%$ lower than the original series. Numerous attempts to train the GAN did not manage to achieve a closer mean value.
CoSMoS got a mean WD score of 275, implying greater distribution similarity, compared to GAN. However, it's DTW score is the worst, and the example plot in Fig. \ref{fig:ud_comparison} displays erratic behavior not found in the original. This can be explained by the autocorrelation structure. 
\afterpage{\clearpage}

\subsection{Air Pollution}
Clearly this data should have non-negative values only, yet the typical values were close to zero. For this reason the non-negative implementation described in section \ref{genmethod} was used here. For Fourier generation, $m=100$ is used. A much higher number of coefficients was kept here, since as visible in Fig. \ref{fig:m_comparison}, even with $m=100$ the synthetic series have a decent amount of deviation from the original. For ARMA, $p=3,q=3$, and in CoSMoS Generalized Gamma dist. was used, with Weibull ACS.

Table \ref{tab:NO_stats} presents the statistical comparison between the methods for the NO data. NO data was characterized by small values around zero, while the time series physically cannot produce negative values. Therefore the non-negative procedure was applied. That is why the Fourier methodology did not conserve the mean and SD in this dataset. In fact both the mean and SD stray the furthest from the original, with deviations of 11.8\% and 11.4\%, respectively. ARMA reproduced the mean and SD. CoSMoS has an avg. mean notably lower than the original, with 6.5\% difference, while it did reproduce the closest skewness values to the original. The kurtosis is especially notable here as it is orders of magnitude greater in this dataset, compared to the others. CoSMoS did the best job on average reproducing the kurtsis. 31\% difference from the original, compared to 106\%  in the next best (GAN). However, the SD is larger than the value itself, indicating large differences in kurtosis between samples.  Fourier has the lowest DTW, but only 5.7\% lower than CoSMoS. CoSMoS got the lowest WD score, slightly lower than GAN.

\begin{figure}[p]
    \centering
    \includegraphics[width=1\linewidth]{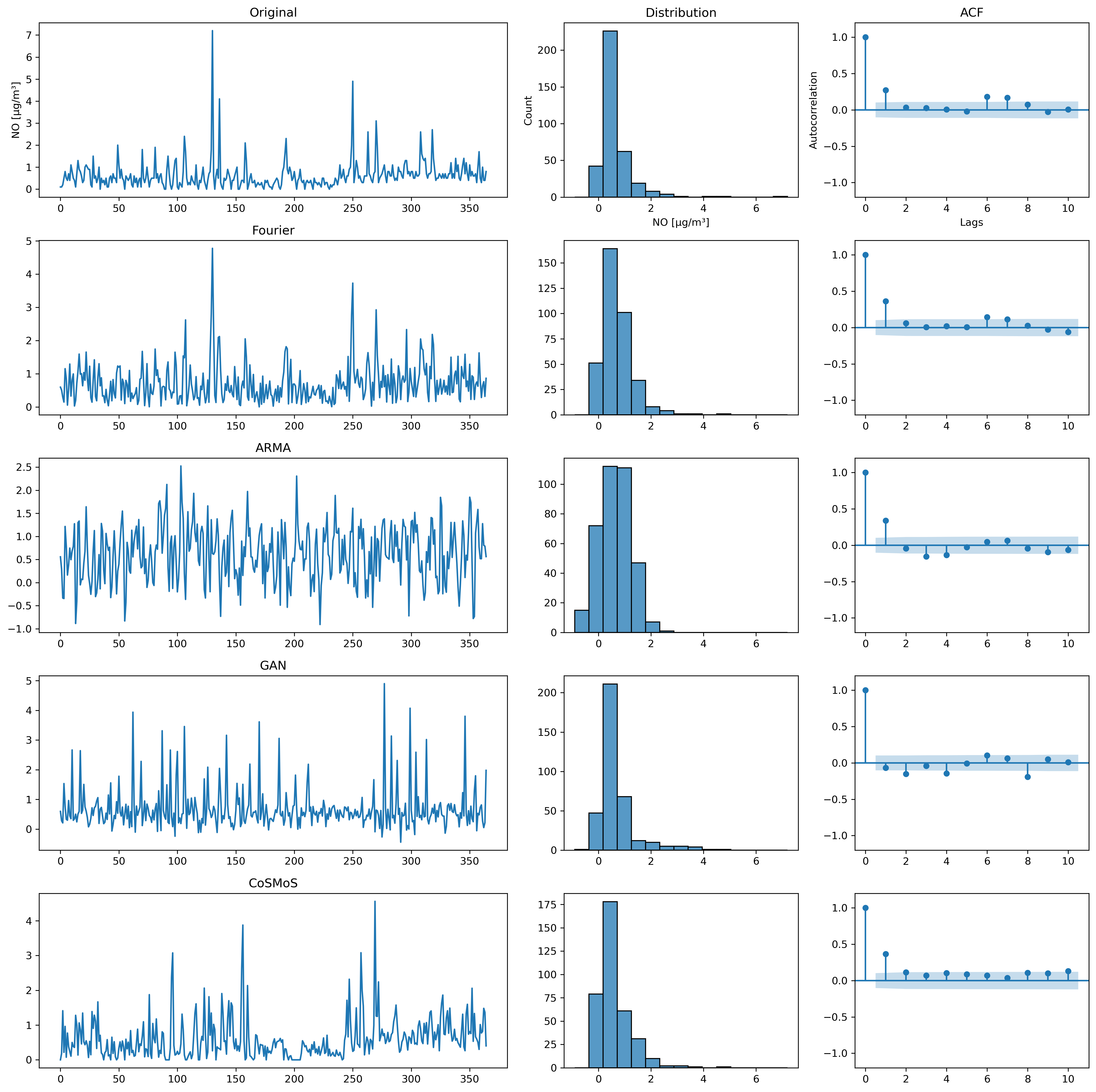}
    \caption{NO - original time series and a single time series produced by each method, including distribution and ACF. }
    \label{fig:NO_comparison}
\end{figure}

A different approach for obtaining non-negative was tested. A synthetic signal was generated, and then any negative values were simply set to zero. This approach, with $m=100$,  yielded a mean of $0.7\pm0.07$, SD of $0.586 \pm 0.009$, skewness of $2 \pm 0.2$, kurtosis of $6.8\pm 1.4$, DTW score of $7.5\pm 1.4$, and WD score of $0.11\pm 0.007$. The averages were close to the original non-negative implementation used. This indicates that this alternative might be also viable in cases with few negative values. 
\begin{table}[ht]
    \centering
        \caption{Statistics and metrics for NO dataset. Average values of the synthetic series $\pm$ SD. Closest statistics and lowest metric scores in bold.
        }
\resizebox{\linewidth}{!}{%
\begin{tabular}{|c|c|c|c|c|c|} 
\hline
\textbf{m=100} & Original & Fourier                  & ARMA                     & GAN               & CoSMoS                      \\ 
\hline
Mean           & 0.64     & 0.72 $\pm$ 0.01          & \textbf{0.64} $\pm$ 0.05 & 0.67 $\pm$ 0.01   & 0.60 $\pm$ 0.04             \\ 
\hline
SD             & 0.65     & 0.58 $\pm$ 0.01          & \textbf{0.65} $\pm$ 0.03 & 0.64 $\pm$ 0.06   & 0.62 $\pm$ 0.13             \\ 
\hline
Skewness       & 4.58     & 1.99 $\pm$ 0.17          & 0.00 $\pm$ 0.13          & 2.83 $\pm$ 0.30   & \textbf{3.52} $\pm$ 1.86    \\ 
\hline
Kurtosis       & 34.25    & 7.09 $\pm$ 1.56          & -0.03 $\pm$ 0.24         & 10.57 $\pm$ 2.82  & \textbf{24.96} $\pm$ 29.32  \\ 
\hline
DTW            & -        & \textbf{7.49} $\pm$ 0.37 & 10.65 $\pm$ 0.35         & 8.84 $\pm$ 0.63   & 7.93 $\pm$ 1.15             \\ 
\hline
WD             & -        & 0.12 $\pm$ 0.01          & 0.26 $\pm$ 0.02          & 0.085 $\pm$ 0.011 & \textbf{0.08} $\pm$ 0.017   \\
\hline
\end{tabular}
}
    \label{tab:NO_stats}
    \end{table}
Despite the fact ARMA reproduced the mean and SD, it has the highest scores in both metrics, as well as averaging a negative kurtosis value. The example series for ARMA in Fig. \ref{fig:NO_comparison} shows negative values are obtained, which is undesirable in this case. The other methods show similarity to the original, with non-negative values in the range $[0,2]$ and occasional sharp peaks. This is also visible in the distribution.

\pagebreak

\subsection{Air Flow Velocity}
$m=5$ is used in Fourier, chosen by trial and error. For ARMA,  $p=2,q=4$, and in CoSMoS Burr type XII was used for both the distribution and the ACS.

\begin{figure}[p]
    \centering
    \includegraphics[width=1\linewidth]{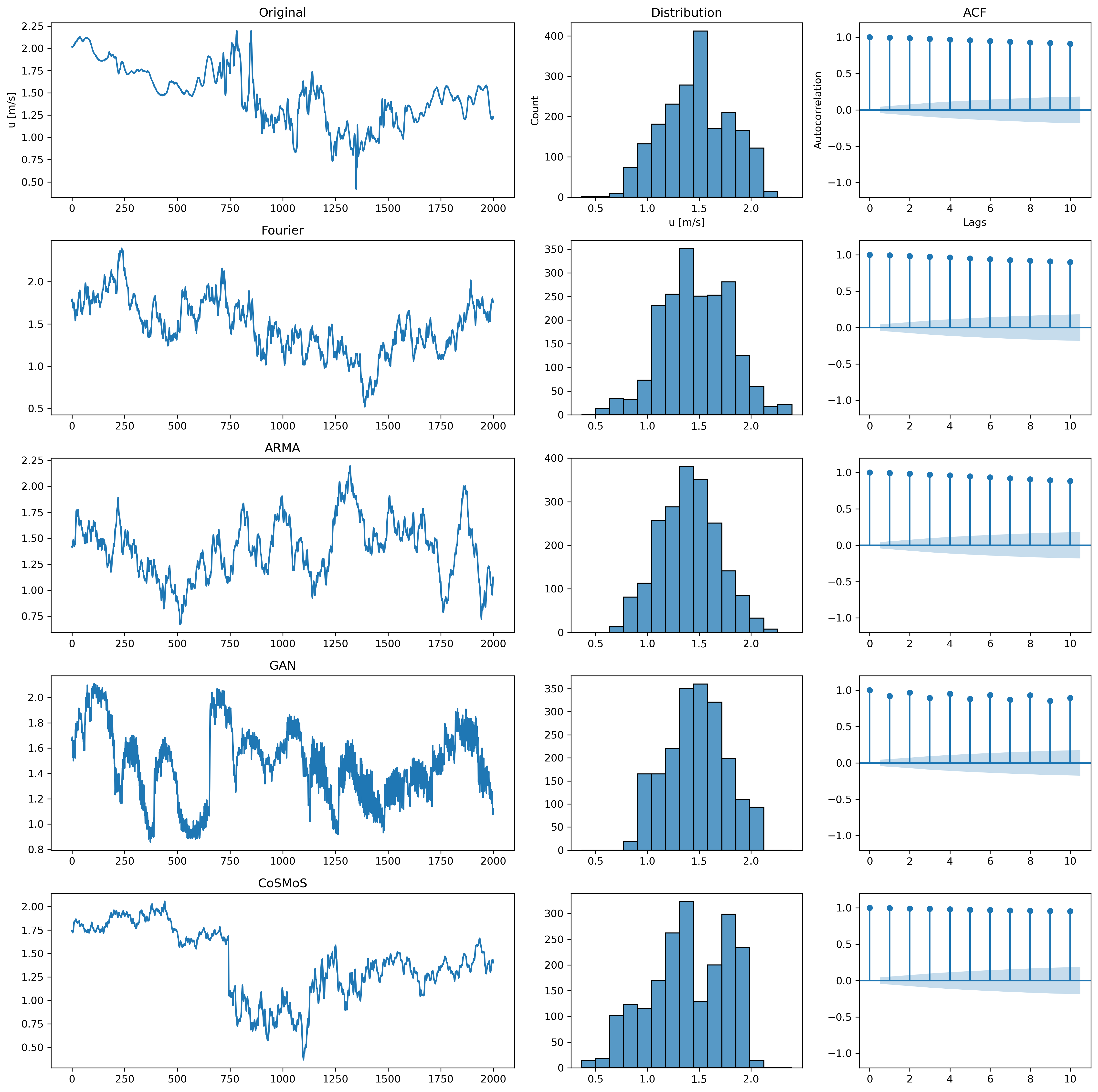}
    \caption{$u$ - The original time series (at the top) and a single time series produced by each method.}
    \label{fig:u_comparison}
\end{figure}

 Looking at table \ref{tab:u_stats}, all methods managed to reproduce the mean with up to 0.01 difference from the original series. Fourier was the only method to reproduce the SD, with CoSMoS coming the closest, with a 10\% lower average. With regards to skewness, ARMA did best. However, the SD of the skewness was 16 times larger than the average value, indicating big variations between different generated series. GAN did best with replicating kurtosis, followed by Fourier. Fourier again scored lowest on both DTW and WD, with CoSMoS being the closest in both. Average DTW for CoSMoS was 11\% more than Fourier, and WD was 103\% greater. GAN and ARMA got similarly high scores in both metrics.
    The GAN example series in Fig. \ref{fig:u_comparison} displays erratic behavior, and ACF values that oscillate rather than decrease monotonically (as in the original). The ARMA sample seems realistic, in terms of the range of values and the distribution and ACF plots. The thing it does not seem to replicate is the general trend of the series itself - in indices up to 1,000 the original signal gets values in the range [1, 2.25], and after it the series presents lower values, in the range of [0.5, 1.75]. The CoSMoS and Fourier samples show similar behavior, but ARMA does not, which might explain the relatively high scores it got from the metrics. 
    
\begin{table}[ht]
    \centering
        \caption{Statistics and metrics for the $u$ dataset (turbulent air flow velocity component). Average values of the synthetic series $\pm$ SD. Closest statistics and lowest metric scores in bold.}
\begin{tabular}{|c|c|c|c|c|c|} 
\hline
\textbf{m=5} & Original & Fourier                    & ARMA                     & GAN                       & CoSMoS             \\ 
\hline
Mean $[m/s]$ & 1.47     & \textbf{1.47} $\pm$ 0      & \textbf{1.47} $\pm$ 0.09 & 1.46 $\pm$ 0.06           & 1.46 $\pm$ 0.06    \\ 
\hline
SD $[m/s]$   & 0.33     & \textbf{0.33} $\pm$ 0      & 0.28 $\pm$ 0.04          & 0.26 $\pm$ 0.03           & 0.30 $\pm$ 0.04    \\ 
\hline
skewness     & 0.003    & 0.07 $\pm$ 0.16            & \textbf{0.02} $\pm$ 0.32 & -0.09 $\pm$ 0.24          & -0.25 $\pm$ 0.31   \\ 
\hline
Kurtosis     & -0.56    & -0.45 $\pm$ 0.26           & -0.18 $\pm$ 0.47         & \textbf{-0.63} $\pm$ 0.31 & -0.41 $\pm$ 0.65   \\ 
\hline
DTW          & -        & \textbf{4.69} $\pm$ 0.47   & 8.79 $\pm$ 2.66          & 8.90 $\pm$ 2.27           & 5.24 $\pm$ 1.04    \\ 
\hline
WD           & -        & \textbf{0.023} $\pm$ 0.005 & 0.089 $\pm$ 0.04& 0.077 $\pm$ 0.03& 0.072 $\pm$ 0.03\\
\hline
\end{tabular}

    \label{tab:u_stats}
    \end{table}

\pagebreak

\subsection{Wave Height}
Here, $m=5$ is used. For ARMA,  $p=6,q=10$, and in CoSMoS Generalized Gamma is used for the distribution, with Burr type XII for the ACS.

  Table \ref{tab:wave_height_stats} details the results for the wave height dataset. As expected, Fourier reproduced the mean and SD. ARMA came the closest with only 2.2\% difference, and GAN the furthest, getting an average mean 26\% smaller than the original. ARMA did the best in terms of reproducing skewness, while CoSMoS did best in kurtosis, with large variability in both. Fourier got the lowest scores in both metrics once again, with ARMA second lowest in both. DTW was 24.3\% higher in ARMA, while WD was 46.5\% higher.
\begin{table}[ht]
    \centering
        \caption{Statistics and metrics for the wave height dataset $\pm$ SD. Closest statistics and lowest metric scores in bold.}
\resizebox{\linewidth}{!}{%
\begin{tabular}{|c|c|c|c|c|c|} 
\hline
\textbf{m=5} & Original & Fourier                    & ARMA               & GAN                & CoSMoS                     \\ 
\hline
Mean         & -0.0092  & \textbf{-0.0092} $\pm$ 0   & -0.009 $\pm$ 0.006 & -0.012 $\pm$ 0.006 & -0.01 $\pm$ 0.018          \\ 
\hline
SD           & 0.04     & \textbf{0.0404} $\pm$ 0    & 0.041 $\pm$ 0.005  & 0.047 $\pm$ 0.004  & 0.031 $\pm$ 0.007          \\ 
\hline
skewness     & -0.17    & \textbf{0.009} $\pm$ 0.27  & -0.0029 $\pm$ 0.31 & -0.079 $\pm$ 0.185 & -0.043 $\pm$ 0.43          \\ 
\hline
Kurtosis     & -0.63    & -0.26 $\pm$ 0.42           & -0.15 $\pm$ 0.51   & -0.95 $\pm$ 0.22   & \textbf{-0.35} $\pm$ 0.58  \\ 
\hline
DTW          & -        & \textbf{0.65} $\pm$ 0.1    & 0.83 $\pm$ 0.13    & 0.86 $\pm$ 0.12    & 1.12 $\pm$ 0.14            \\ 
\hline
WD           & -        & \textbf{0.005} $\pm$ 0.002 & 0.008 $\pm$ 0.003  & 0.009 $\pm$ 0.004  & 0.018 $\pm$ 0.009          \\
\hline
\end{tabular}
}
\label{tab:wave_height_stats}
\end{table}

\begin{figure}[p]
    \centering
    \includegraphics[width=1\linewidth]{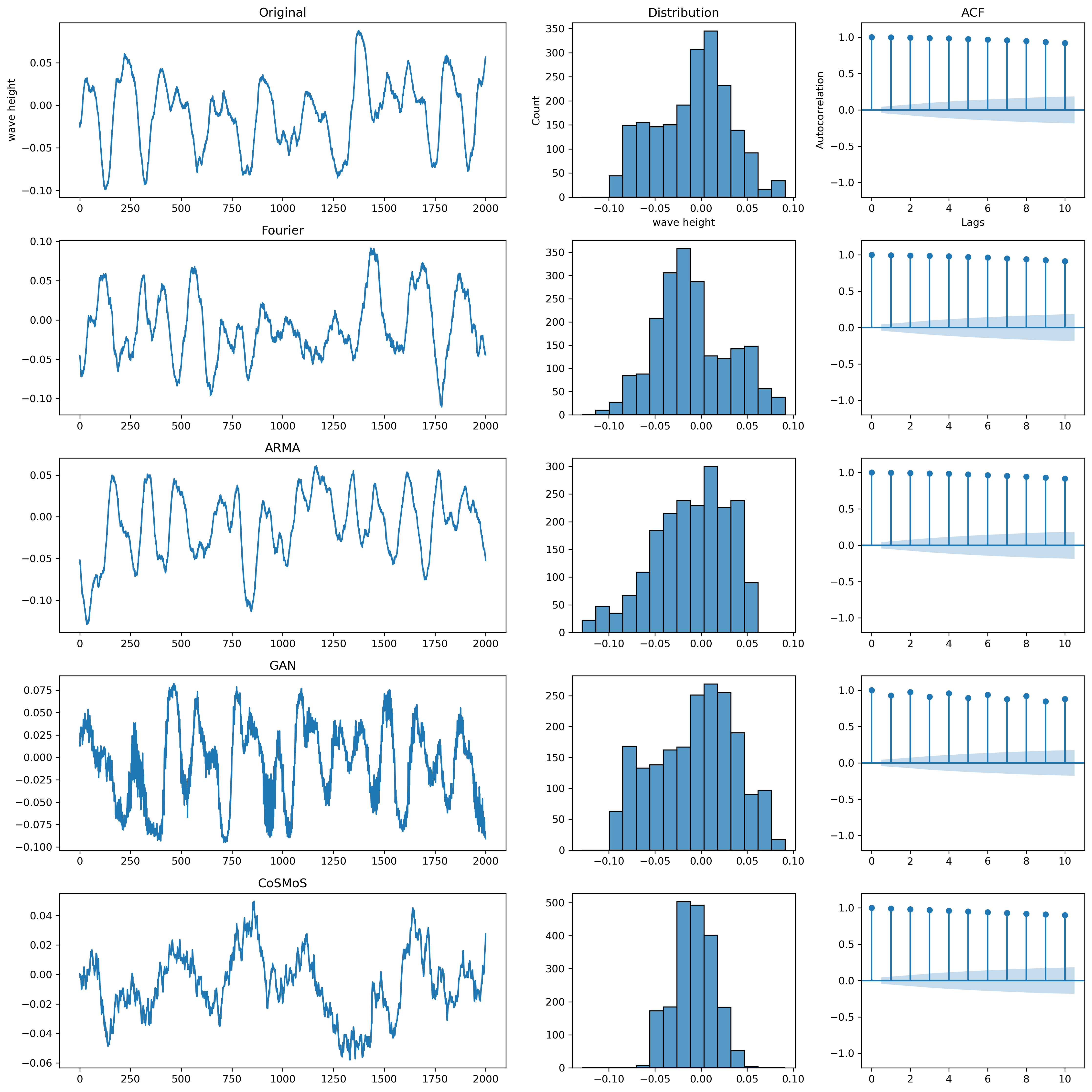}
    \caption{Wave Height - The original time series (at the top) and a single time series produced by each method.}
    \label{fig:wave_height_comparison}
\end{figure}

    Fig.   \ref{fig:wave_height_comparison} gives a similar assessment as to the performance of the methods. Fourier and ARMA have samples closely resembling the original. In GAN, the general trends and the distribution appear similar, but the series values change much more sharply than in the original. Also, similar to the $u$ sample, the ACF oscillates in a way not present in the original, at least not to a visible degree. Note how the CoSMoS sample receives a much smaller range of values than the original and others, as evident in the distribution plot. 

\pagebreak

\section{Conclusions}
In all datasets, Fourier did a superior job in reproducing both the mean and SD, and got the lowest DTW and WD scores. This shows the capability of this method to take a single sample and produce thousands of different synthetic time series with similar statistical properties, visual shape, autocorrelation, and distribution. The ability of the method to set the similarity level allows for a great flexibility that is not available in all other methods. The method is also simple to implement and computationally efficient, which make it scalable. The various examples clearly show that the method is also versatile, and robust to the temporal resolution of the original signal.

A straightforward direction for future research is to expand this idea to higher dimensions, using the multi-dimensional Fourier transform. Another possibility is to use this method for data augmentation of a machine-learning algorithm - i.e. to use this method to generate synthetic series for training a GAN. Finally, this method currently uses only a single instance of time series data. If, for example, the same set of data from different years is available, this method may be further developed so it could leverage this additional information to produce a wider range of synthetic series.

\section{Acknowledgements}
We wish to thank Prof. Dan Liberzon for the wind and waves data shared for this paper. 

\section{Software Availability}
The code and datasets used in this paper are shared in an online repository (code will be made available upon acceptance).

\bibliography{STSG}

\end{document}